\documentclass[preprint]{aastex62}

\usepackage{verbatim}
\usepackage{apjfonts}
\usepackage{ulem}
\usepackage{mathptmx}
\usepackage{natbib}
\usepackage{soul}
\usepackage{graphicx}
\usepackage{xcolor}
\usepackage{hyperref}
\usepackage{booktabs}
\usepackage{nccmath}
\usepackage{epstopdf}

\newcommand{\chandra}{ {\it Chandra} }

\newcommand{\spitzer}{ {\it Spitzer} }

\newcommand{\mum}{$\,\mu$m}

\newcommand{\fpa}{\color{black}}
\newcommand{\sasha}{\color{black} }
\newcommand{\rick}{\color{black}}

\newcommand{\nico}{\color{black}}
\newcommand{\sashaa}{\color{black} }
\newcommand{\rickk}{\color{black}}

\begin{document} 

\title{\sasha{SPECTRAL PROPERTIES OF POPULATIONS BEHIND THE COHERENCE \\ IN  $\spitzer$ near-infrared and $\chandra$ X-RAY BACKGROUNDS}}
\author{Yanxia Li}
\affiliation{Institute for Astronomy, University of Hawaii at Manoa, Honolulu, HI 96822, USA}
\author[0000-0002-1697-186X]{{Nico Cappelluti}}
\affiliation{Physics Department, University of Miami, Coral Gables, FL 33124, USA}
\author[0000-0002-0797-0646]{{G\"{u}nther Hasinger}}
\affiliation{Institute for Astronomy, University of Hawaii at Manoa, Honolulu, HI 96822, USA}
\affiliation{European Space Astronomy Centre (ESA/ESAC), Director of Science, E-28691 Villanueva de la Ca\~nada, Madrid, Spain} 
 \author[0000-0001-8403-8548]{{Richard G. Arendt}}
\affiliation{Observational Cosmology Laboratory, Code 665, Goddard Space Flight Center, 8800 Greenbelt Road, Greenbelt, MD 20771, USA}
\affiliation{CRESST II / University of Maryland, Baltimore County, 1000 Hilltop Circle, Baltimore, MD 21250, USA}
\author{{Alexander Kashlinsky}}
\affiliation{Observational Cosmology Laboratory, Code 665, Goddard Space Flight Center, 8800 Greenbelt Road, Greenbelt, MD 20771, USA}
\affiliation{SSAI, Lanham, MD 20706, USA}
\author[0000-0001-9879-7780]{Fabio Pacucci}
\affiliation{Department of Physics, Yale University, New Haven, CT 06511, USA}
\affiliation{Kapteyn Astronomical Institute, Groningen, 9747 AD, Netherlands}
\correspondingauthor{Nico Cappelluti}
\email{ncappelluti@miami.edu}


\begin{abstract}

We study the coherence of the near-infrared and X-ray background fluctuations and the X-ray spectral properties of the sources producing it. We use  data from multiple $\spitzer$ and $\chandra$ surveys, including the UDS/SXDF surveys, the Hubble Deep Field North, the EGS/AEGIS field, the $\chandra$ Deep Field South and the COSMOS surveys, comprising $\sim$2275 $\spitzer$/IRAC hours and $\sim$~16 Ms of $\chandra$ data collected over a total area of $\sim$~1~deg$^2$. We report an overall $\sim$5$\sigma$ detection of a cross-power signal on large angular scales $>$ 20$''$ between the 3.6 and 4.5\mum\ and the X-ray bands, with the IR vs [1-2] keV signal detected at 5.2$\sigma$.  The [0.5-1] and [2-4] keV bands are correlated with the infrared wavelengths at a $\sim$1$-$3$\sigma$ significance level. The hardest X-ray band ([4-7] keV) alone is not significantly correlated with any infrared wavelengths due to poor photon and sampling statistics. We study the X-ray SED of the cross-power signal. We find that its shape is consistent with a variety
of source populations of accreting compact objects, such as local unabsorbed AGNs
or high-z absorbed sources. We cannot exclude that the 
excess fluctuations are produced by more than one population.
Because of poor statistics, the current relatively broad photometric bands employed here do not allow distinguishing the exact nature of these compact objects or if a fraction of the fluctuations have instead a local origin.

\end{abstract}

\keywords{cosmology: observations --- infrared: diffuse background --- quasars: supermassive black holes --- X-rays: diffuse background}



\section{Introduction}\label{s:intro}

Sources presently undetected at any wavelength leave footprints in cosmic backgrounds. In the near-infrared (IR) photons from the cosmic infrared background (CIB) reveal otherwise undetected star forming galaxies, AGNs, the Galaxy and even objects from the reionization epoch that, because of redshift, may peak in the {\em Spitzer} bands.

{\rickk Measuring the absolute levels of the CIB suffers from large systematic uncertainties associated with foreground subtraction. Studying CIB fluctuations is a promising alternative, because it is much less sensitive to the absolute zero-point of the measurements.} {\sasha In addition, a fluctuation analysis could potentially  distinguish undetected  components, e.g. early black holes, (BHs) and intra halo light (IHL), from known populations  of stars, galaxies and AGN \citep{Kashlinsky96}. This opens up the possibility of isolating high-$­z$ emissions due to the distinct spectral amplitude and structure of the underlying early sources \citep{Cooray04,Kashlinsky04}}.
 
There have been a few studies about the coherence between the cosmic X-ray background (CXB) and the CIB \citep{Cappelluti13,Mitchell-Wynne16, Cappelluti17c,Li18}. 
These studies mostly focused on the cross power spectra, which were used to study the possible contributing sources to the cross power. {\fpa The observed coherence between CIB and CXB has been explained by BHs, e.g., direct collapse BHs \citep[DCBH,][]{Yue13}, which are high-mass (10$^{4-6}$ M$_\odot$) black hole seeds predicted to form in pristine, high-$z$ environments \citep{Bromm03}, or accreting primordial BHs \citep[PBH,][]{WpaperPBH,Kashlinsky16}, which are stellar-mass objects formed during or after the inflation period. }{\sashaa Following these theoretical proposals, \citet{Kashlinsky05b,Kashlinsky07b} identified in deep Spitzer-based analysis significant source-subtracted CIB fluctuations at levels much higher than expected from remaining known sources \citep{Kashlinsky12,Helgason12a,Helgason14}. See the recent review by \citet{Kashlinsky18} for summary and discussion.}

 In addition to studying the cross-power spectrum itself, it is also possible to use the SED
of the mean cross powers as a distinguishing diagnostic of the X-ray spectra of the
different source models. In fact, \citet{Li18} first studied the SED shape of the cross-power signal using COSMOS data only, but found that they cannot effectively discriminate various models due to limitations of the data in their study. 
\cite{ Cappelluti13,Cappelluti17c} showed that   while the coherence between the CIB and CXB is of the order 10-20\%, this is a strong function of the angular scales, approaching 1 between 20-1000$\arcsec$. Moreover,   the coherence between the CIB and the CXB in different sub-bands is essentially the same \citep[See][]{Li18}.  In this  paper we are interested in characterizing the excess power in the 20\arcsec-1000\arcsec angular range,  and given the considerations listed above we can reasonably assume that  the populations responsible for this signal in the CIB versus CXB cross power are intrinsically highly  correlated. This might suggest that the same source populations produce the observed excess large scale cross-power. For this reason we can assume that using the amplitude of the cross-power is a tool to estimate the stacked (or average) X-ray spectrum of the unknown sources producing the excess fluctuations. The shape of such a SED can help investigate the nature of these sources. 
The goal of this paper is to provide the spectrum of these correlated sources, although this work does not provide a definitive distinction between the several source populations
proposed to produce this signal.

{\nico In this work, for the first time, we combine the data from multiple fields with deep $\spitzer$ and $\chandra$ data} and probe the SED of the CIB-CXB cross power signal at 4 narrow X-ray bands: [0.5-1] keV,  [1-2] keV,  [2-4] keV,  and [4-7] keV, in order to study the X-ray spectral properties of the CIB-CXB cross power. 


\section{Data sets and Map Making}\label{s:data}

\subsection{Chandra X-ray data }\label{s:xdata}
We collect and analyze the X-ray data from 5 surveys: the $\chandra$ Legacy Survey of the UKIDSS Ultra Deep Survey Field \citep[UDS,][]{Kocevski18}, the Hubble Deep Field North \citep[HDFN,][]{Alexander03}, the $\chandra$ ACIS-I AEGIS survey \citep[EGS, ][]{Goulding12}, the $\chandra$ Deep Field South \citep[CDFS,][]{Luo17} and the $\chandra$ COSMOS-Legacy Survey \citep[COSMOS,][]{Civano16}. Some information about each field is listed in Table~\ref{tab:table1}.   These are 5 of the deepest and most studied fields by both {\em Chandra} and {\em Spitzer}. The choice of these fields is also driven by
their deep coverage with HST to allow future cross-correlation studies with shorter wavelength optical and NIR data. 
Following \citet{Li18}, we reduce our data using the Chandra Interactive Analysis of Observations software \citep[CIAO, ][]{Fruscione06} with the main procedures summarized as below. The level=1 event files are re-calibrated and cross-matched with corresponding optical catalogs to further improve the absolute pointing accuracy. We examine the light curves of the background and clean the flares which otherwise could contaminate the real cosmic background signal. The exposure maps are evaluated at a single energy value (middle value of each band) to avoid introducing bias from modeling. Similar procedures are applied on the stowed data, which were taken when the ACIS detector was out of the focal plane, and these data are used in estimating the particle-only background \citep{Hickox06}. 

We number the X-ray photons sequentially by their arrival time and map them into images ``A'' (using odd numbered photons) and ``B'' (using even numbered photons). A and B maps have the same exposure time and have been observed simultaneously so that the difference map of the two (A-B) only contains instrumental effects. We then create the mosaic signal maps (A+B) and noise maps (A-B) to estimate the noise level.

The mosaic maps are created in 4 narrow bands: [0.5-1] keV, [1-2] keV, [2-4] keV and [4-7] keV. This combination of bands
is the result of a tradoff between increasing the number of bands, and maximizing the S/N per band.
The X-ray masks are created by removing any pixels within a 7$''$ radius around each point source in the corresponding source catalogs (Table~\ref{tab:table1}). {\fpa The mask radius is chosen to remove $>$90\% brightness of the detected X-ray sources \citep{Civano16}}. The extended emissions from groups and clusters of galaxies are identified and incorporated in the masks as well \citep{Finoguenov10,Erfanianfar13,Erfanianfar14,Finoguenov15}. 

The X-ray maps are matched to the IR maps (Section~\ref{s:irdata}), so that they have the same astrometry and pixel scales. 
The CDFS and HDFN have smaller pixel scales (i.e. 0.6\arcsec vs. 1.2\arcsec) because the deeper coverage of these fields facilitates subpixelization of the data which allows slightly more detailed source masking. No changes in the large scale power have been noted between these choices of pixel scale of the final images. 
Combining the X-ray masks with the corresponding IR masks, we obtain the final masks ($M_{\rm IR, X}$) for the CIB-CXB cross power analysis. 
 As shown by \citet{Helgason14}, IR maps are deeper than X-ray maps for typical X-ray sources. This means that even if the depth of the surveys is different in the X-ray the band, the maps have very similar shot noise levels as shown by comparing   Figs. 14 and 17 of \citet{Li18}.
The numbers of photons before and after masking in each band are shown in Table~\ref{tab:table1}  together with flux limits of the surveys in the [0.5-2] keV band.

\subsection{Spitzer IR data }\label{s:irdata}
Our IR data (3.6 and 4.5\mum) are taken with $\spitzer$/IRAC from multiple programs: the Spitzer Extended Deep Survey program \citep[UDS: program ID = 61041, EGS: program ID = 61042,][]{Fazio11}, the GOODS Legacy program \citep[HDFN: program ID = 169, CDFS: program ID = 194,][]{Dickinson03}, and the $\spitzer$ Large Area Survey with the HyperSuprime-Cam \citep[COSMOS: program ID = 90042,][]{Steinhardt14}. 

Detailed data reduction is described in \citet[][and references therein]{Li18}, with the main procedures summarized as below. The reduction starts with the corrected basic calibrated data (cBCD). The data are grouped and processed with the self-calibration method \citep{Arendt10}, with the model described as $\rm \it D^i = S^{\alpha}+F^{p}+F^{q} $, where $D^i$ is the measured intensity of a single pixel $i$ from the single frame $q$, and $S^\alpha$ is the true sky intensity at location $\alpha$. $F^p$ and $F^q$ describe the offset between the observed and the expected sky intensity, where $F^p$ remains constant with time and records the offset for detector pixel $p$, while $F^q$ is variable during the observations. The self-calibration helps to remove some artifacts\footnote{\href{http://irsa.ipac.caltech.edu/data/SPITZER/docs/irac/iracinstrumenthandbook}{http://irsa.ipac.caltech.edu/data/SPITZER/docs/irac/iracinstrumenthandbook}} on the images from, e.g., {\fpa``first-frame effect'', which is due to the fact that the detector response is noticeably different before and after long slews when the detector is periodically scanning the field during the course of the observations}. 

Each field has 2 mosaic maps at both IR wavelengths created by merging the calibrated frames together. Source detection and masking are done on the mosaic maps with 2 major steps: (1) source modeling, to identify and remove point sources and resolved extended sources; (2) masking (sigma clipping), to clean up the artifacts from modeling and the remaining emission from bright sources \citep{Arendt10}. 

 The source-subtracted CIB fluctuations could be contributed by the shot noise from remaining sources in the beam, and the clustering of the remaining CIB sources. The CIB fluctuations are measured at a given shot noise level, $P_{\rm SN}=\int$ S$^2(m) dN(m)$, which we express in units of ~nJy~nW~m$^{-2}$sr$^{-1}$ \citep[see][ Sect IV.A.1]{Kashlinsky18}. The COSMOS maps are clipped to the shot noise level of $\sim$86 ~nJy~nW~m$^{-2}$sr$^{-1}$ and other maps are at the level of $\sim$50 and $\sim$30~nJy~nW~m$^{-2}$sr$^{-1}$ in 3.6 and 4.5\mum, respectively.


\section{Fourier analysis of the fluctuation maps}\label{s:fft}

For each field, at each IR and X-ray wavelength, we have the mosaic maps with resolved sources masked out by applying $M_{\rm IR,X}$. We then use the Fourier transforms (via an FFT algorithm) to extract the cross-power spectrum for each possible pair of the IR and X-ray wavelengths. 

Following the conventions in \citet{Kashlinsky12} and \citet{Cappelluti13}, the fluctuation map is defined as:
\begin{equation}
    \delta F({\bf x})=  I({\bf x})  -  \langle I({\bf x})\rangle,
\end{equation}
 and its 2D Fourier transform is 
 \begin{equation}
    \Delta ({\bf q}) = \frac{1}{4\pi^2}\int\delta F({\bf x})exp(-i{\bf x} \cdot {\bf q} ) d^2x. 
 \end{equation}
 
 The auto-power spectrum (at a single band) is 
 \begin{equation}
 P(q) =\langle \mid \Delta({\bf q}) \mid ^2 \rangle, 
 \end{equation}
 averaged over $[q,q+\delta q]$. The error estimation of $P(q)$ is,
 \begin{equation}
\sigma_{P(q)} = P(q)/\sqrt{0.5N_q},
 \end{equation}
 where 0.5$N_{\rm q}$ is the number of independent Fourier elements. 
The CIB-CXB cross powers are calculated using 
\begin{equation}
  P_{1\times2}(q) =\langle\Delta_{1}(q)\Delta^{*}_{2}(q)\rangle,
\end{equation}
 with the errors 
 \begin{equation}
       \sigma_{P_{1\times2}(q)}  =\sqrt{P_1(q)P_2(q)/N_q}. 
 \end{equation}

Throughout the paper, we also compute the mean squared fluctuations, which are calculated as $q^2P(q)/2\pi$ as a function of the angular scales, 2$\pi$/$q$. 
In order to combine signals, all of the maps from different fields have the same Fourier binning to give power at identical angular frequencies ($q$).



 \section{Results}\label{s:results}
\subsection{Cross-Power Spectra}\label{s:irx}
 
\begin{figure*}
\centering
        \includegraphics[trim=0 380 0 0, clip,width=1\textwidth]{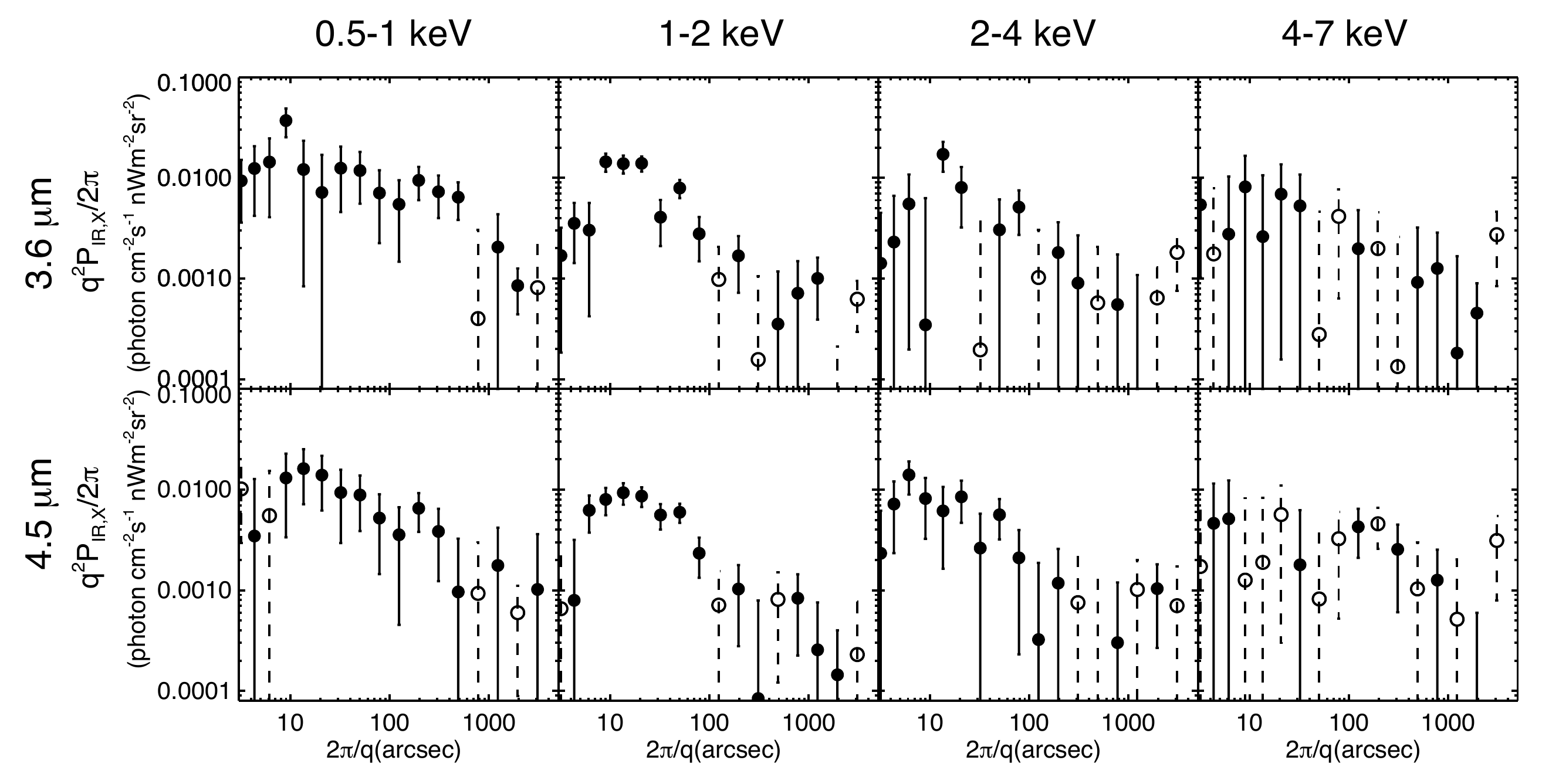}
    \caption{Stacked CIB-CXB cross-power spectra: between 3.6 \mum\ and [0.5-1] keV, [1-2] keV, [2-4] keV and [4-7] keV (top panels); between 4.5 \mum\ and CXB (bottom panels). Open circles with dashed error bars denote the absolute values of negative results. Note that the shown spectra are subtracted by the cross power between IR and the stowed X-ray map, that is, the potential contamination from the X-ray particle background is removed from our final results. The final mean powers (used for further SED analysis) are summarized in Table~\ref{tab:table2}. 
        \label{fig:irx} }
\end{figure*}

\begin{figure*}
\centering
        \includegraphics[width=1\textwidth, angle=90]{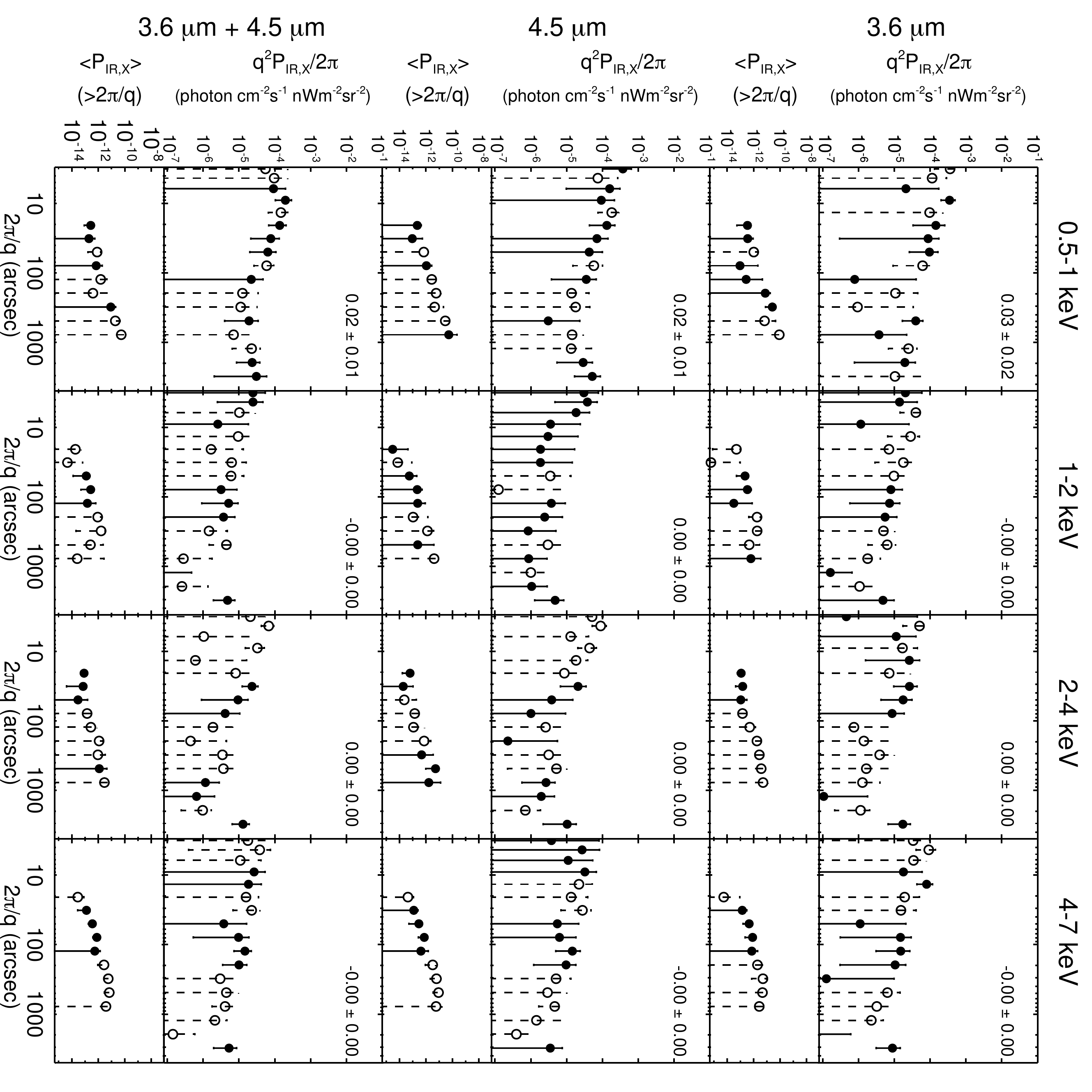}
    \caption{Stacked CIB-Stowed background  cross-power spectra: between 3.6 \mum\ and [0.5-1] keV, [1-2] keV, [2-4] keV and [4-7] keV  stowed background (upper panels); between 4.5 \mum\ and stowed background (middle panels). Open circles with dashed error bars denote the absolute values of negative results. For completeness we show also the combined 3.6$\mu$m+4.5$\mu$m vs stowed background in the (bottom panels). In the labels we show the average cross-power above 20$\arcsec$.
        \label{fig:irxres} }
\end{figure*}

{\rickk The stacked cross powers are calculated by the simple mean of the powers from the 5 subfields, instead of weighted averages. Because uncertainties of the cross powers are derived from the signal, randomly low (or high) measurements are assigned artificially small (or large) uncertainties. Therefore, the weighted averages will be biased towards the lower values and the overall uncertainties could be biased. }  We calculate the cross powers between IR maps and the stowed X-ray images (reprojected to the  astrometric frame of the IR images) in the same manner and find that the cross powers are consistent with zero for all pairs  as shown in Figure \ref{fig:irxres}. The cross powers with the stowed X-ray maps are subtracted from the cross-power spectra. The resulting cross powers are shown in Figure ~\ref{fig:irx}. 
 
  As demonstrated in \cite{ Cappelluti13,Cappelluti17c, Mitchell-Wynne16, Li18} the CIB vs CXB cross-power is in excess over shot noise at angular scales larger than 20$\arcsec$. Since at those scales known foregrounds are significantly weaker than our signal \citep{Cappelluti17c,Kashlinsky18,Helgason14}, therefore we calculate the weighted mean cross power above 20$''$  as 
 \begin{equation}
      \langle q^2 P_{\rm IR,X}\rangle/\langle q^2 \rangle,
 \end{equation}
 
 between each IR and X-ray band, for each field and for the stacked cross-power spectrum.  As shown in Table~\ref{tab:table2}, the [1-2] keV band is significantly correlated with all IR wavelengths at $\sim$5$\sigma$ significance level, while the [0.5-1] and [2-4] keV bands are correlated with the IR wavelengths at $\sim$1-3$\sigma$ levels. There is no significant cross power signal between the [4-7] keV band and any IR wavelengths so  we cannot exclude an intrinsic correlation. By including the COSMOS field with a larger area, our new analysis extends the CIB-CXB cross power analysis to $\sim$3000$''$, doubling the largest scale reached in \citet{Cappelluti17c}. 
  The reason the highest significance is in the [1-2] keV band is due the higher signal-to-noise ratio due to the peak in effective area of the {\em Chandra} mirror in that spectral region. Stacked results in the hard X-ray band (i.e. [2-4] keV + [4+7]  keV) are not as significant as in \citet{Li18} since deep fields add data affected by strong cosmic variance  on very large scales. In the hard band most of the events are particles, so the low astrophysical hard X-ray photon statistics is another source of noise. 
 As a cross-check we compared the significance of our stacks with those shown in Fig. 17 of \citet{Li18} for the [2-7] keV band vs 3.6 $\mu$m and 4.5 $\mu$m and found consistent results.
 
\subsection{Cross-Power Hardness Ratios}\label{s:hr}
Similar to the X-ray color-color diagram, we look into three hardness ratios for the mean cross power, which are equivalent of photometric color indices, following the typical definition \citep{Brunner08}:
\begin{equation}
HR1 = \frac{\langle P_{\rm IR,~[1-2] keV}\rangle - \langle P_{\rm IR, ~[0.5-1] keV} \rangle}{\langle P_{\rm IR, ~[1-2] keV}\rangle + \langle P_{\rm IR, ~[0.5-1] keV}\rangle}, 
\end{equation}
\begin{equation}
HR2 = \frac{\langle P_{\rm IR, ~[2-4] keV}\rangle - \langle P_{\rm IR, ~[1-2] keV} \rangle}{\langle P_{\rm IR,~ [2-4] keV}\rangle + \langle P_{\rm IR, ~[1-2] keV}\rangle},
\end{equation}
\begin{equation}
HR3 = \frac{\langle P_{\rm IR, ~[4-7] keV}\rangle - \langle P_{\rm IR, ~[2-4] keV} \rangle}{\langle P_{\rm IR,~ [4-7] keV}\rangle + \langle P_{\rm IR,~ [2-4] keV}\rangle}, 
\end{equation}

The statistical 1$\sigma$ hardness ratio errors are calculated from the mean power (Table~\ref{tab:table2}) by error propagation. The mean 1$\sigma$ errors for the hardness ratios are $\sim$0.3 for $HR1$ and $HR2$ and $\sim$0.8 for $HR3$. Due to the larger errors in $HR3$, resulting from the uncertainties from harder bands, we only examine the relations between $HR1$ and $HR2$.

\begin{figure*}
\centering
        \includegraphics[trim=0 20 0 10, clip,width=0.8\textwidth]{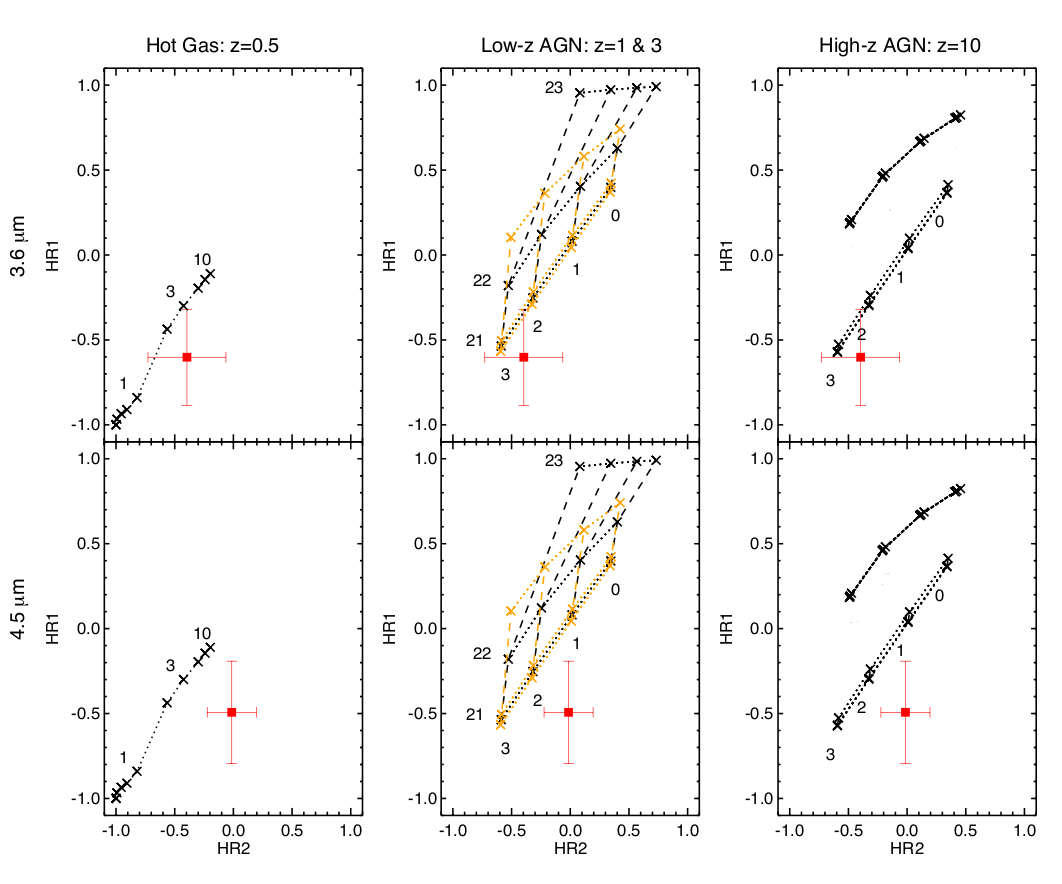}
 \caption{ Color-color plot or the hardness ratios (HR1 vs. HR2) of $\langle P_{IR, X}\rangle$, with the top row showing HRs using $\langle P_{3.6{\mu}m, X}\rangle$ and the bottom row using $\langle P_{4.5{\mu}m, X}\rangle$. The data point with $\langle P_{IR, X}\rangle$ calculated above 20$''$ in each panel (summarized in Table~\ref{tab:table2}) is highlighted with red squares. The error bars represent the 1 sigma uncertainty on the HR
values. We present 3 cases of the theoretical models for: hot gas (left column), AGNs at $z$=1 \& 3 (middle column) and high-z AGNs (right column). The track in the left column shows the Bremsstrahlung spectrum with $kT$ ranging from 0.05 to 10 keV, labels represent the gas temperature. In the middle column, grid lines show the locations of colors for different absorbed power-law spectra. Dashed lines connect spectra with spectral index $\Gamma$ = 0 to 3 as labeled. Dotted lines connect spectra with column densities from log(N$_H$) = 21 to log(N$_H$) = 23 as labeled on the left side of the curves. The grid for $z$ = 3 is highlighted in orange. In the right column, the upper track (dashed-line) is from the {\sc pexrav} model (i.e. a Compton thick AGN with N$_{\rm H}$=1.5$\times$10$^{24}$~cm$^{-2}$), with leaking flux ({\sc rel\_refl}) ranging from 1\% to 30\% of the total flux, marked from left to right, respectively. The lower double-dotted track is from the simple power-law model at high-$z$ with spectral index  $\Gamma$ = 0 to 3 as labeled plus galactic absorption. See Section \ref{s:hr} for more details about the models. \label{fig:hr}}
\end{figure*}

{\rickk Figure~\ref{fig:hr} shows plots of the observed hardness ratios in comparison to various models, with the grid lines referring to single-component models for the X-ray emission. } We examine 3 cases:

(1) Hot gas emissions ($z$ = 0.5):  a thermal plasma model with Galactic absorption, with the model defined as {\sc wabs*apec} in {\sc Xspec} notation \citep[version 12.10.0,][]{Arnaud96}. ``{\sc apec}'' is an emission spectrum from collisionally-ionized diffuse gas calculated from the AtomDB atomic database\footnote{\href{ http://atomdb.org}{ http://atomdb.org}}. The temperature $T$ ranges from $kT =  0.05$ keV to 10 keV.

\begin{figure*}
\centering
        \includegraphics[trim=0 20 0 0, clip,width=0.9\textwidth]{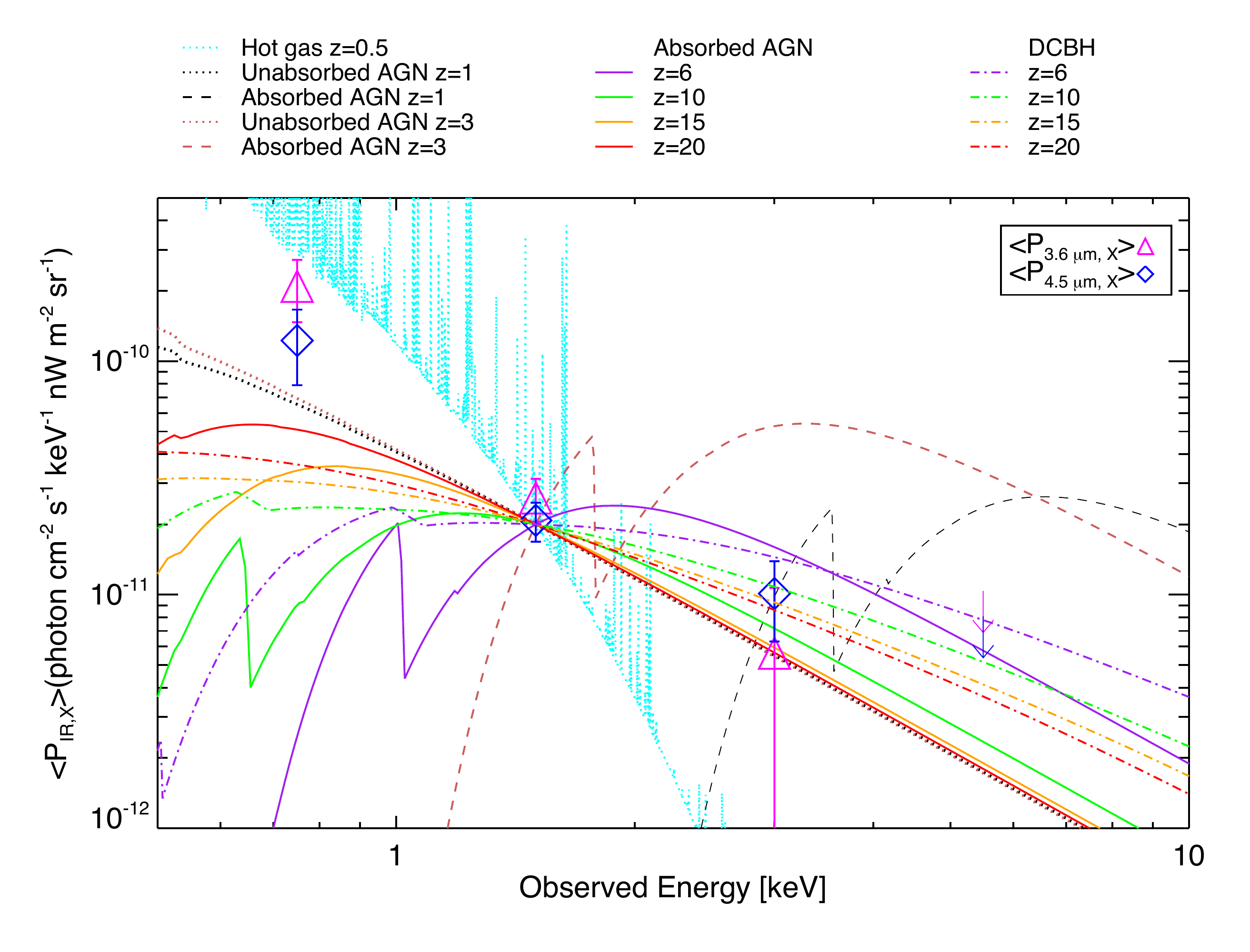}
    \caption{The mean cross power (calculated above 20$''$) between 3.6 \mum~(magenta triangles) or 4.5\mum~(blue diamonds) and narrow X-ray bands of [0.5-1] keV, [1-2] keV, [2-4] keV, [4-7] keV as a function of the observed X-ray energy (0.75 keV, 1.5 keV, 3 keV, and 5.5 keV). The 2$\sigma$ upper limit of the measurement  at 5.5 keV are shown as downward arrows. We also show the X-ray spectral models of low-$z$ sources: $z=3$ AGNs (dark red), $z$=1 AGNs (black) and hot gas (cyan). The absorbed AGNs at $z$ of 6, 10, 15, 20 are shown in purple, green, orange and red solid lines, while the DCBH models with dash-dot lines (Section~\ref{s:sed}).   \label{fig:sed_model} }
\end{figure*}

{\rick Although a hot gas model with $kT$ $\sim$ 3$-$10~keV  at z=0.5 might match the data as shown in Figure~\ref{fig:hr}, it is not physically possible. Such a high temperature probably corresponds to a massive cluster of galaxies, e.g. 10$^{14}$M$_{\odot}$. {\sashaa By construction, these clusters are actually masked out and their source density is expected to be $<$1 in our field of view. \citet{Cappelluti13} has also found that removing X-ray clusters with an extra X-ray mask does not alter the CIB fluctuations, so clusters do not have significant contributions.} If the observed X-ray emission is due to hot gas, the temperature should be significantly lower, as adopted in Figure~\ref{fig:sed_model}}. It is worth noticing that an analysis of the XB\"ootes field by \citet{Kolodzig17} showed instead that the bulk of large scale fluctuations  in the maps was produced by galaxy clusters and groups in the mass range  10$^{13-14}$M$_{\odot}$. However, such an analysis was obtained with an average exposure of 5 ks over the field of view. Our stacked results are obtained with an average exposure of $\sim$ 2 Ms. Our observations  are on average 400 times deeper than theirs, so the diffuse sources  producing the large scale fluctuations in their observations have been masked here. Therefore we can safely exclude moderately massive clusters as contributors to the cross power signal.  
 
(2) A possible alternative is that a missing population of intermediate redshift AGN  (at $z=1-3$)  could be responsible for the excess. In order to model such a population we approximate its spectrum with a simple absorbed power-law model that accounted for the Galactic hydrogen absorption and a possible intrinsic absorption at the source redshift, with the model defined as {\sc wabs*zwabs*zpo}. Each grid line in~Figure~\ref{fig:hr} (middle column) corresponds to this model spectrum with photon indices $\Gamma$ = 0, 1, 2, 3 and intrinsic absorption (in the observer frame) of the Galactic hydrogen column density (N$_H$) = 10$^{21}$, 10$^{22}$ and 10$^{23}$~cm$^{-2}$. 
 By carefully examining  Figure~\ref{fig:hr}, we find that the hardness ratios strongly disfavor a scenario where the bulk of the population is produced by a moderate redshift highly-obscured AGN population. Our data are instead consistent with a weakly absorbed  power-law with a spectral index of the order $\Gamma\sim$2-3, which is  rather steep for type 1 AGNs (see, e.g. \citealt{Just07}). 
The mean colors are very soft, and therefore if an absorbed AGN population is present, it would also require an additional softer component which may be associated with star formation.

(3)  It has been postulated that a possible population of very high-$z$ AGNs ($z$=10) in the form of DCBHs could be responsible for the large-scale cross power. These sources are expected to be extremely Compton thick and show two main peaks in the SED, one at  $\sim$2-5 $\mu$m and another  at 1 keV in the observers' frame. These SEDs have been presented in several works (e.g., \citealt{Yue13,Pacucci15, Pacucci17, Pacucci18}). The X-ray components of these SEDs are shown in Fig. \ref{fig:sed_model} folded through the {\em Chandra} response matrices.
However, we also wanted to test the case for un-absorbed high-$z$ AGN, so we computed hardness ratios with with 2 types of models. One is a single power-law with intrinsic absorption ({\sc zwabs*zpo}, with $\Gamma$ = 0, 1, 2, 3 and column density N$_H$ = 10$^{21}$, 10$^{22}$ and 10$^{23}$~cm$^{-2}$). The other one is for special Compton-thick sources, whose spectra are dominated by a Compton-reflection continuum from a cold medium, which could be produced by the inner side of the putative obscured torus plus a soft power-law and is made by the photons ``leaking'' through the absorber \citep{Brunner08}. To reproduce such a spectrum and to compute the expected hardness ratios, we use the {\sc Xspec} model {\sc pexrav},  i.e., an exponentially cutoff power-law spectrum reflected from neutral material \citep{Magdziarz95} in a highly obscured environment (N$_{\rm H}$=1.5$\times$10$^{24}$~cm$^{-2}$), with leaking flux ({\sc rel\_refl}) ranging from 1\% to 30\% of the total flux observed. The data appear to be better explained by the single power-law model, although the spectra must be harder than for local AGNs.

\subsection{Cross-Power SED modeling}\label{s:sed2}

In Figure~\ref{fig:sed_model}, we present the SED of the mean cross power. For comparison, we also present X-ray spectral models of absorbed and unabsorbed AGNs, hot gas, and DCBHs. The models for DCBHs are drawn from the radiation-hydrodynamic simulations presented in \citet{Pacucci15}. These simulations require as inputs: 1) initial mass of the seed and 2) initial mass distribution and metallicity of the host galaxy.
A typical choice for the initial mass of a DCBH is $\sim 10^5 \, \mathrm{M_{\odot}}$, following also the study  on the initial mass function of intermediate mass black holes by \citet[][]{Ferrara14}. As the mass increases (within the typical mass range of DCBHs) the normalization of the spectrum increases, with minimal modifications to the overall shape. We can thus interpret the spectral models presented in this paper as typical spectral shapes for DCBHs with initial mass $\sim 10^5 \, \mathrm{M_{\odot}}$. When it comes to the time evolution, typical evolutionary times for these systems are of order 100 Myr. As long as the metallicity and column density do not vary dramatically during the time evolution, the spectral shape undergoes minimal variations, mostly recognizable as modifications of the spectral lines.

The modeled spectra in Fig. 4 are rescaled to match the mean levels of the measurements at 1-2 keV.   We point out that a quantitative statistical interpretation of the SED is not possible with the current data quality. Hence, the comparison of data with models in this section is  qualitative rather than quantitative.
{\rickk In order to examine the implications of the SEDs independent of the results from the hardness ratios, we do not limit our parameter selections to the results from Section~\ref{s:hr}}. For AGNs, we assume a simple standard power-law model with two absorption components ({\sc wabs*zwabs*zpo} at redshift of $z$~=~1,3). The first component models the Galactic absorption with a fixed $N_{\rm H}$ of 1.72$\times$10$^{20}$ cm$^{-2}$. The second component represents the AGN intrinsic absorption. We choose common values for the power-law photon index ($\Gamma$~=~1.9) and $N_{\rm H}$ (1.5$\times$10$^{24}$ cm$^{-2}$ for absorbed AGNs, 10$^{21}$cm$^{-2}$ for unabsorbed AGNs, see e.g. \citealt{Caccianiga04}). For the hot gas (Bremsstrahlung spectrum), we use {\sc Xspec} model ({\sc wabs*apec}) with $kT$ = 0.5~keV, at $z$~=~0.5. {\fpa The predictions for the SEDs of AGNs and DCBHs are computed in the high redshift domain, at $z$ = 6, 10, 15, 20. }{\sasha The SEDs of the DCBH models specifically tuned to the measured CIB fluctuations can also be found in \citet{Yue13}.} 

\subsection{Data  versus SED model comparisons}\label{s:sed}

Interpreting the cross powers, $<$P$\_{IR,X}>$, shown in Fig. 4 as the SED of the correlated X-ray background is consistent (within uncertainties) whether the 3.6 or 4.5 micron cross power is examined. 
By comparing the measurements to the models  both using the SEDs and hardness ratios, we rule out the possibilities of low-$z$ absorbed AGN and hot  $kT>$3 keV gas for reproducing our measurements.
The other models at high-$z$ (absorbed AGNs and DCBHs) appear to fit the data but they cannot be differentiated due to the negligible difference in their shapes.  The differences between absorbed AGNs and DCBHs visible in Figure~\ref{fig:sed_model} are mostly due to the presence of metal features in the former class of spectra, while in the latter class the assumption is that the absorbing gas is pristine.   There is an indication of an excess at the softest [0.5-1] keV band, which cannot be fully accounted for by any of the models. This might suggest that at softest energy a thermal Galactic component might correlate with residual Galactic cirrus. 
The most likely spectral shape for the correlating components is either a steep power-law or an AGN-like power law plus a very soft thermal component. 
Our main conclusion is that by using spectral analysis alone we can affirm that, above 1 keV, the spectrum of primary sources producing CXB fluctuations correlating with the CIB is a power-law, which consistent with radiation produced by accretion. At lower energies  we see an indication of an excess signal over such a power-law so we cannot exclude a thermal origin of  the spectrum pointing to either a Galactic component or to  very low mass virialized sources (i.e. groups). In fact \cite{Cappelluti13} showed that the main contributors to the large scale CXB in the CDFS are galaxy groups.
However the contribution of these sources to the CIB is still
the subject of investigation, since while emission from IHL \citep{Cooray12} has been
proposed as a major contributor to the CIB it is not clear if IHL can account for the cross
correlation with the X-rays.
 On the other hand \citet{Cappelluti17a} proposed that a possible contributor to the cross-power is  scattering of X-ray light on galaxy dust but  a recent work by \citet{Ricarte19} shows that this effect is subdominant. 
Therefore, through spectral analysis we have evidence that at least part  of the  sources producing our detected signal are compact accreting objects. However, spectral information is not detailed enough to determine their redshift and detailed population properties.



\section{Summary}\label{s:discussion}
 
In order to explain the measured cross power between CIB and CXB fluctuations, we analyze currently available datasets from multiple fields and make calculations with stacking techniques. We find that the CIB is most significantly correlated with the [1-2] keV background at a $\sim$5$\sigma$ level.  Other bands are correlated at lower significance, but 
overall we detect a signal whose significance stays constantly  above 5$\sigma$ when adding harder energy bands to the [0.5-1] keV.

In addition, mimicking the studies of X-ray colors using the count rates in different X-ray bands, we visualize the hardness ratios using the mean cross power. We find that a power-law model, consistent with local accreting  unabsorbed AGN or high-z absorbed AGN are favored if the sources are accreting compact objects.  Local absorbed AGN and hot $kT>$3 keV  gas are not consistent with our data. 
 Lastly, we compare the measured SED of the cross powers with various models and we find that unabsorbed AGNs, and BHs (AGN and DCBH) at high $z$ ($>$15) can explain the measurements better than other scenarios. We see a deviation from the power-law model at very soft energies suggesting a possible Galactic contamination of the signal.
However, with only 4 X-ray bands over a large range of energies, it is difficult to distinguish what kind of compact sources is producing the signal.

Further improvements in the data quality, especially at even larger angular scales from both the IR and the X-ray \citep{kerosita, WPaperCIB} are needed to explain the origin of the excess cross-power fluctuations. Some other techniques, e.g. Lyman tomography \citet{Kashlinsky15a}, could help to pin down the exact redshift of the source populations.

\begin{table*}
  \centering
\caption{X-Ray Map Properties}
  \label{tab:table1}
  \begin{tabular}{llllll}
\toprule
    			&        UDS &      HDFN &       EGS &      CDFS &    COSMOS    \\
 \midrule
\# of pointings  &  25 & 20 & 97 & 101 & 117 \\
 Area	 (deg$^2$)		 &       0.12 &      0.02 &      0.10 &      0.04 &      0.74   \\
 Pixel Scale ($''$) 	&        1.2 &       0.6 &       1.2 &       0.6 &       1.2   \\  
 S$_{lim,0.5-2~keV}$ (cgs) & 1.4$\times$10$^{-16}$ & 2.5$\times$10$^{-17}$ &  2.0$\times$10$^{-16}$ & 6.4$\times$10$^{-18}$ & 2.2$\times$10$^{-16}$ \\
 \%$_{mask}$ 	&      37.8\% &     37.5\% &     31.8\% &     39.1\% &     46.6\%    \\
 N$_{\rm ph, [0.5-1]} ^{\rm u}$ &     41551 &    134065 &     56547 &    710949 &     75360     \\
 N$_{\rm ph, [0.5-1]}^{\rm m}$ &     24084 &     65207 &     30539 &    356632 &     31628     \\
N$_{\rm ph, [1-2]}^{\rm u}$ &     71955 &    209295 &    114557 &   1475705 &    170025     \\
N$_{\rm ph, [1-2]}^{\rm m}$ &     36172 &     84905 &     58264 &    646442 &     62753     \\
N$_{\rm ph, [2-4]}^{\rm u}$ &    100203 &    255677 &    168682 &   2114792 &    235235   \\
N$_{\rm ph, [2-4]}^{\rm m}$ &     56227 &    136167 &    105535 &   1151228 &    110308    \\
N$_{\rm ph, [4-7]}^{\rm u}$ &    132249 &    302467 &    192454 &   2420475 &    264213    \\
N$_{\rm ph, [4-7]}^{\rm m}$ &     78629 &    176220 &    126445 &   1406109 &    133298    \\
\bottomrule
\end{tabular}

      \small
      \vspace{0.2cm}
      Notes: We summarize the X-ray photon counts before masking (N$_{\rm ph}$$^{\rm u}$) and after masking (N$_{\rm ph}$$^{\rm m}$) for each X-ray narrow band, in each field. References for the used catalogs are as follows. UDS: \citet{Kocevski18}, HDFN: \citet{Alexander03}, EGS: \citet{Goulding12}, CDFS: \citet{Luo17}, COSMOS: \citet{Civano16}  \\
   
\end{table*}

    \begin{table*}[h!]
  \centering
  \caption{Cross-Power Spectrum Amplitude ($>$20$''$), calculated as $ \langle q^2P_{IR, X}\rangle/\langle q^2 \rangle $\footnote{in units of 10$^{-11}$ photon s$^{-1}$ cm$^{-2}$ nW m$^{-2}$ sr$^{-1}$}}
  \label{tab:table2}
  \begin{tabular}{llllll}
    \toprule
    &  & [0.5-1] keV & [1-2] keV & [2-4] keV & [4-7] keV \\
    \midrule

        & 3.6 \mum &               1.17 $\pm$    15.61 ( 0.1 $\sigma$)  &               2.07 $\pm$     1.77 ( 1.2 $\sigma$)  &              -1.55 $\pm$     2.80 (-0.6 $\sigma$)  &              -2.98 $\pm$     4.11 (-0.7 $\sigma$)  \\
UDS        & 4.5 \mum &               2.90 $\pm$     9.91 ( 0.3 $\sigma$)  &               1.94 $\pm$     1.06 ( 1.8 $\sigma$)  &               1.41 $\pm$     1.74 ( 0.8 $\sigma$)  &               2.29 $\pm$     2.64 ( 0.9 $\sigma$)  \\
        & 3.6 \mum &               3.91 $\pm$     3.59 ( 1.1 $\sigma$)  &               1.53 $\pm$     1.28 ( 1.2 $\sigma$)  &              -2.54 $\pm$     2.56 (-1.0 $\sigma$)  &              -1.21 $\pm$     3.61 (-0.3 $\sigma$) \\
HDFN        & 4.5 \mum &               1.60 $\pm$     2.85 ( 0.6 $\sigma$)  &               1.50 $\pm$     1.02 ( 1.5 $\sigma$)  &              -2.61 $\pm$     2.06 (-1.3 $\sigma$)  &              -1.97 $\pm$     2.86 (-0.7 $\sigma$)  \\
        & 3.6 \mum &               6.07 $\pm$     3.31 ( 1.8 $\sigma$)  &               2.37 $\pm$     1.00 ( 2.4 $\sigma$)  &               1.56 $\pm$     2.12 ( 0.7 $\sigma$)  &               4.01 $\pm$     2.99 ( 1.3 $\sigma$)  \\
EGS        & 4.5 \mum &               8.57 $\pm$     2.47 ( 3.5 $\sigma$)  &               3.24 $\pm$     0.74 ( 4.4 $\sigma$)  &               1.44 $\pm$     1.56 ( 0.9 $\sigma$)  &               0.11 $\pm$     2.21 ( 0.0 $\sigma$)  \\
        & 3.6 \mum &               5.35 $\pm$     2.51 ( 2.1 $\sigma$)  &               3.74 $\pm$     0.61 ( 6.1 $\sigma$)  &               2.54 $\pm$     1.18 ( 2.2 $\sigma$)  &               1.12 $\pm$     1.58 ( 0.7 $\sigma$)  \\
CDFS        & 4.5 \mum &               4.44 $\pm$     2.30 ( 1.9 $\sigma$)  &               2.26 $\pm$     0.58 ( 3.9 $\sigma$)  &               1.88 $\pm$     1.05 ( 1.8 $\sigma$)  &              -3.27 $\pm$     1.38 (-2.4 $\sigma$)  \\
        & 3.6 \mum &              20.03 $\pm$     4.86 ( 4.1 $\sigma$)  &               2.87 $\pm$     1.06 ( 2.7 $\sigma$)  &               2.63 $\pm$     2.06 ( 1.3 $\sigma$)  &               1.99 $\pm$     2.93 ( 0.7 $\sigma$)  \\
COSMOS        & 4.5 \mum &               8.91 $\pm$     3.76 ( 2.4 $\sigma$)  &               1.78 $\pm$     0.82 ( 2.2 $\sigma$)  &               5.19 $\pm$     1.60 ( 3.3 $\sigma$)  &               2.69 $\pm$     2.26 ( 1.2 $\sigma$)  \\
\hline
        & 3.6 \mum &              10.45 $\pm$     3.12 ( 3.3 $\sigma$)  &               {\bf 2.60 $\pm$     0.54 ( 4.8 $\sigma$)}  &               1.12 $\pm$     1.02 ( 1.1 $\sigma$)  &               0.21 $\pm$     1.45 ( 0.1 $\sigma$)  \\
 STACK & 4.5 \mum &                6.13 $\pm$     2.18 ( 2.8 $\sigma$)  &              {\bf  2.08 $\pm$     0.40 ( 5.2 $\sigma$)}  &               2.02 $\pm$     0.76 ( 2.7 $\sigma$)  &               0.25 $\pm$     1.09 ( 0.2 $\sigma$)  \\

     \bottomrule
  \end{tabular}
  \end{table*}



\vspace*{0.2cm} 
\acknowledgments
The authors thank the anonymous referee for the valuable suggestions and for the time spent improving our work. Support for this work was provided in part by NASA through ADAP grant NNX16AF29G. N.C. acknowledges $\chandra$ SAO grant AR6-17017B and AR4-15015B. F.P. acknowledges support from $\chandra$ SAO grant AR8-19021A and from the NOVA Fellowship. NASA's support for the Euclid LIBRAE project NNN12AA01C is gratefully acknowledged. 

\newpage

\end{document}